# INFLUENCE OF CARBON CONTENT ON THE CRYSTALLOGRAPHIC STRUCTURE OF BORON CARBIDE FILMS

O. Conde[1a], A. J. Silvestre[2] and J. C. Oliveira[1b]

[1] Departamento de Física, Faculdade de Ciências, Universidade de Lisboa, Campo Grande, Ed. C1, 1749-016 Lisboa, Portugal.
[2] Instituto Superior de Transportes, R. Castilho nº 3, 1269-074 Lisboa, Portugal.

## Abstract

Boron carbide thin films were synthesised by laser-assisted chemical vapour deposition (LCVD), using a $CO_2$ laser beam and boron trichloride and methane as precursors. Boron and carbon contents were measured by electron probe microanalysis (EPMA). Microstructural analysis was carried out by Raman microspectroscopy and glancing incidence X-ray diffraction (GIXRD) was used to study the crystallographic structure and to determine the lattice parameters of the polycrystalline films. The rhombohedral-hexagonal boron carbide crystal lattice constants were plotted as a function of the carbon content, and the non-linear observed behaviour is interpreted on the basis of the complex structure of boron carbide.

## Introduction

The rhombohedral boron carbide, often denoted $B_4C$ since his composition was established by Ridgway in 1934 [1], is the most stable compound in the boron-carbon system and exists as a single-phase material over a wide range of solubility, generally accepted from about 9 to about 20 at% C.

From the technological point of view, boron carbide exhibits many attractive properties, such as low specific weight and high hardness, even surpassing diamond and cubic boron nitride at temperatures over 1100 ºC [2]. Moreover, it presents high melting point (2450 ºC) and modulus of elasticity, and a great resistance to chemical attack. These properties make boron carbide an interesting wear and corrosion-resistant ceramic material for thin film applications. Besides, the $B_4C$ has been utilised as a neutron absorbent material in the nuclear industry since it has a high neutron capture cross-section and is also a good candidate for high temperature thermoelectric energy converters due to its high temperature stability [3].

The confluence of all these properties in one unique material reflects the complex structure of boron carbide, for which some uncertainty still persists [4,5]. The most accepted model for the rhombohedral structure of boron carbide considers $B_{11}C$ icosahedra clusters directly linked by covalent bonds and indirectly linked by a C-B-C chain along the main diagonal of the rhombohedron [5−7]. Based on this model, boron carbide has a $B_{12}C_3$ stoichiometry with 20 at% C. The wide range for carbon content characteristic of the boron carbide is made possible by the substitution of boron and carbon atoms for one another within both $B_{11}C$ icosahedra and C-B-C chains.

[a] Corresponding author: Tel: +351 1 7500035, Fax: +351 1 7573619, E-mail: oconde@fc.ul.pt
[b] Present address: Faculdade de Ciências e Tecnologia, 3030 Coimbra

The aim of this paper is to present experimental results on the microstructure, chemical composition, deposition rate and structure of boron carbide films deposited by laser-assisted chemical vapour deposition (LCVD), and to discuss the influence of carbon content on the crystallographic structure of the deposited films.

## Experimental procedure

Boron carbide films were deposited on silica substrates using a $CO_2$ laser beam as heat source and a dynamic reactive gas mixture of $BCl_3$, $CH_4$ and $H_2$. The experimental apparatus used for the experiments was already described in reference [8].

The $CO_2$ laser was operated in cw $TEM_{00}$ mode at a wavelength of 10.6 μm and impinges the substrate surface at perpendicular incidence, with a diameter of 13 mm. No focusing lens was used since fused silica substrates absorb approximately 84 % of the laser radiation. The reactor was pumped to a base pressure less than $2x10^{-6}$ Torr before the introduction of the reaction gas mixture which consists of $BCl_3$ (purity 99.99%), $CH_4$ (purity 99.9995%), $H_2$ (purity 99.9995%) and Ar (purity 99.9995%). Mass flow controllers regulate the reactant flow rates while the total pressure was measured by a capacitance manometer and kept constant by means of a throttle valve. Fused silica plates with $15x15x2$ $mm^3$ were used as substrates. They were ultrasonically cleaned in acetone and ethanol prior to insertion in the reactor.

Since the substrates were always kept stationary under the laser irradiation, the experimental variables of the set-up are the laser output power (P), the interaction time ($t_{int}$), the total pressure ($p_t$) and the partial flow rates of each gas ($\varphi_i$). In this study, total pressure, hydrogen flux and argon flux were kept constant at 100 Torr ($1.33x10^4$ Pa), 200 sccm and 430 sccm, respectively. The other experimental parameters were varied in the ranges listed in table 1.

Table 1
Process parameters for LCVD of boron carbide films

| *Experimental parameters* | *Range of values* |
|---|---|
| Laser output power (W) | 125 – 175 |
| Interaction time (s) | 30 – 90 |
| Partial flux of $BCl_3$ (sccm) | 34 – 40 |
| Partial flux of $CH_4$ (sccm) | 7.2 – 14 |

The relative amount of the carbon and boron precursors in the reactive atmosphere can be characterised by the parameter $\phi = \varphi_{CH4}/(\varphi_{CH4} + \varphi_{BCl3})$. In this study, $\phi$ took values between 0.15 and 0.29, which were optimised to achieve homogeneous deposits of $B_4C$, as reported previously [8,9].

Thickness profiles were obtained by optical profilometry and the microstructure was examined by scanning electron microscopy (SEM). Quantitative chemical analysis was performed by electron probe microanalysis (EPMA). Structural analyses were carried out with a micro-Raman spectrometer ($Ar^+$ laser, 488 nm excitation line) and glancing incidence X-ray diffraction (GIXRD, glancing angle of 1º) using Cuk$\alpha$ radiation.

## Results and discussion

### *Morphology and microstructure*

According to visual observation, all the deposited films are homogeneous, exhibit the characteristic $B_4C$ light grey shining colour and good adherence. Since the laser beam has a nearly gaussian energy profile the deposits are approximately circular in shape, with diameters between 5.5 and 7.4 mm. Results of EPMA revealed that the coatings are chemically homogeneous, showing evidence of carbon concentration profiles almost flat with distance from spot centre, and that the stoichiometry of the films (9 – 20 at% C) is mainly determined by gas phase composition [9].
SEM analyses of different films have shown that their morphology is nodular with a fine and uniform grain structure all over the deposits. A typical microstructure is illustrated in figure 1a where a SEM micrograph of a film deposited with P = 150 W, $t_{int}$ = 30 s and $\phi$ = 0.25 is shown. The film presents a polycrystalline structure that seems to have been originated from a high density of nucleation centres yielding a mean grain size of few tenths of micron. The micrograph also shows some cracks formed during the cooling of the samples, due to the difference in the thermal expansion coefficients of the film ($4.5\times10^{-6}$ $K^{-1}$) and the silica substrate ($0.5\times10^{-6}$ $K^{-1}$).

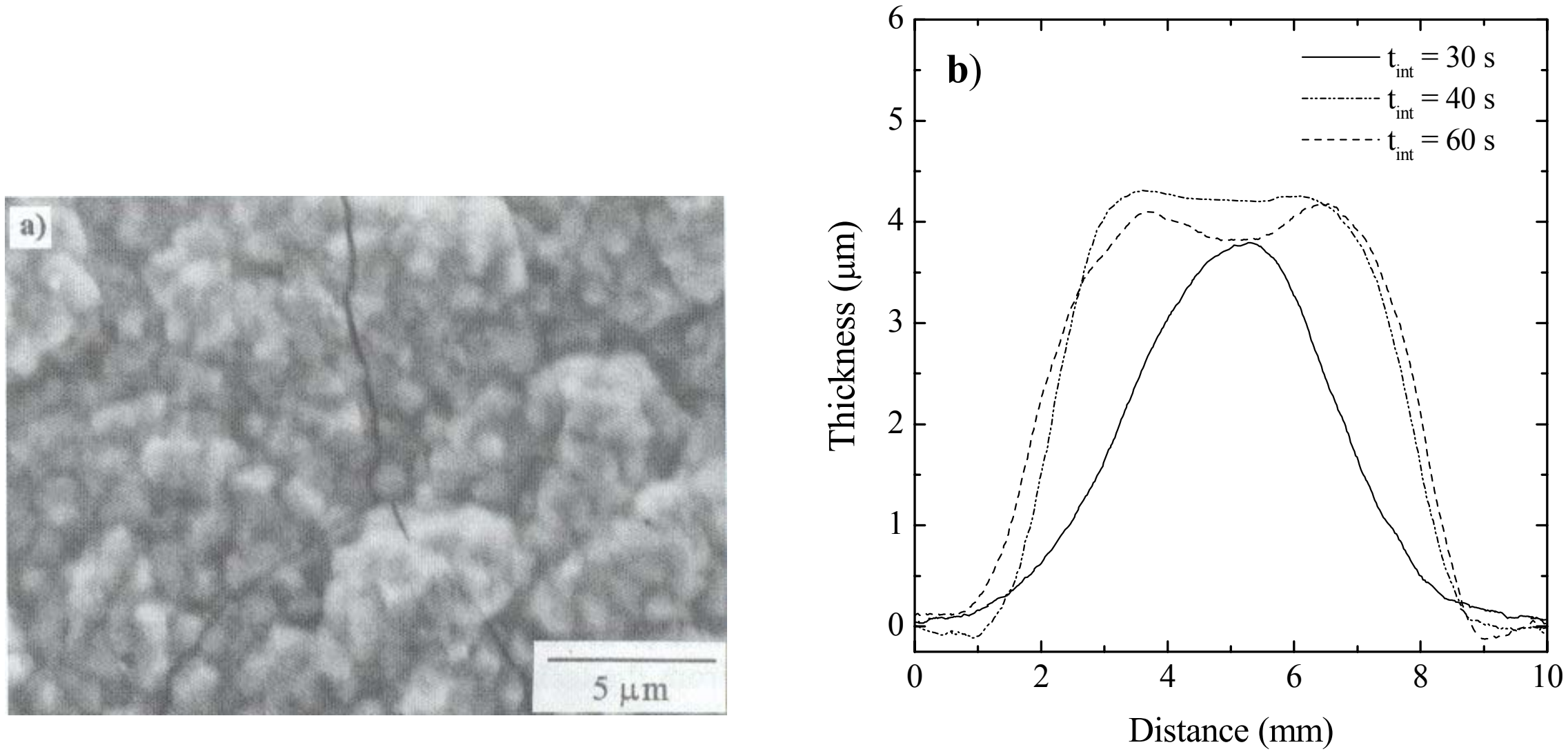

**Figure 1** - a) SEM micrograph of a film deposited with $\phi$ = 0.25, P = 150 W and $t_{int}$ = 30 s; b) Thickness profiles of films deposited with $\phi$ = 0.25, P = 150 W and different interaction times.

Concerning the topography of the deposited material, three different thickness distributions were found, namely gaussian, top-flat and volcano-like profiles. Figure 1b shows thickness profiles measured on spots deposited with $\phi$ = 0.25, P = 150 W and different interaction times. As can be seen, for $t_{int}$ = 30 s the spot presents a broad-gaussian profile; by increasing the irradiation time up to 40 s the thickness distribution of the deposited material evolves to a quasi top-flat profile. The film processed with $t_{int}$ = 60 s presents a central depression associated with a lack of film material yielding to a volcano-like profile. The series of coating profiles plotted in figure 1b also illustrates an important observed behaviour: for constant ϕ and P values, the thickness of the films, defined as the maximum height above the substrate surface, does not show any significant dependence on the laser irradiation time. However, the amount of deposited material is an increasing function of the irradiation time, as can be determined from the increasing area of the region limited by the thickness profiles.

*Growth kinetics*

Thickness profiles were very useful to investigate the growth kinetics of the $B_4C$ films, providing not only a mean to estimate the total amount of deposited material but also the values of the surface temperature induced by laser-material interaction. The apparent deposition rate for the boron carbide films was evaluated in terms of the total mass deposited per unit time, following the method described in ref. [10] and assuming the bulk density of $B_4C$, 2.52 g.cm$^{-3}$. Apparent deposition rate values in the range 2.3 to 9.4 µg.s$^{-1}$ were obtained. Following the calculation technique described in ref. [9], the surface temperature achieved at the centre of the films during deposition was estimated between 800 K and 1080 K.
Figure 2 shows the Arrhenius diagram where the apparent deposition rates and temperatures are those referred above. The fact that only one straight line is needed to fit the data indicates that only one mechanism is responsible for $B_4C$ film growth. From the slope of the straight line the apparent activation energy was calculated to be 27.5±7.8 kJ.mol$^{-1}$. The low magnitude of the activation energy indicates a small dependence of the deposition rate on temperature, which means that mass transport of reactive gaseous species is the rate-limiting step for the boron carbide film growth.

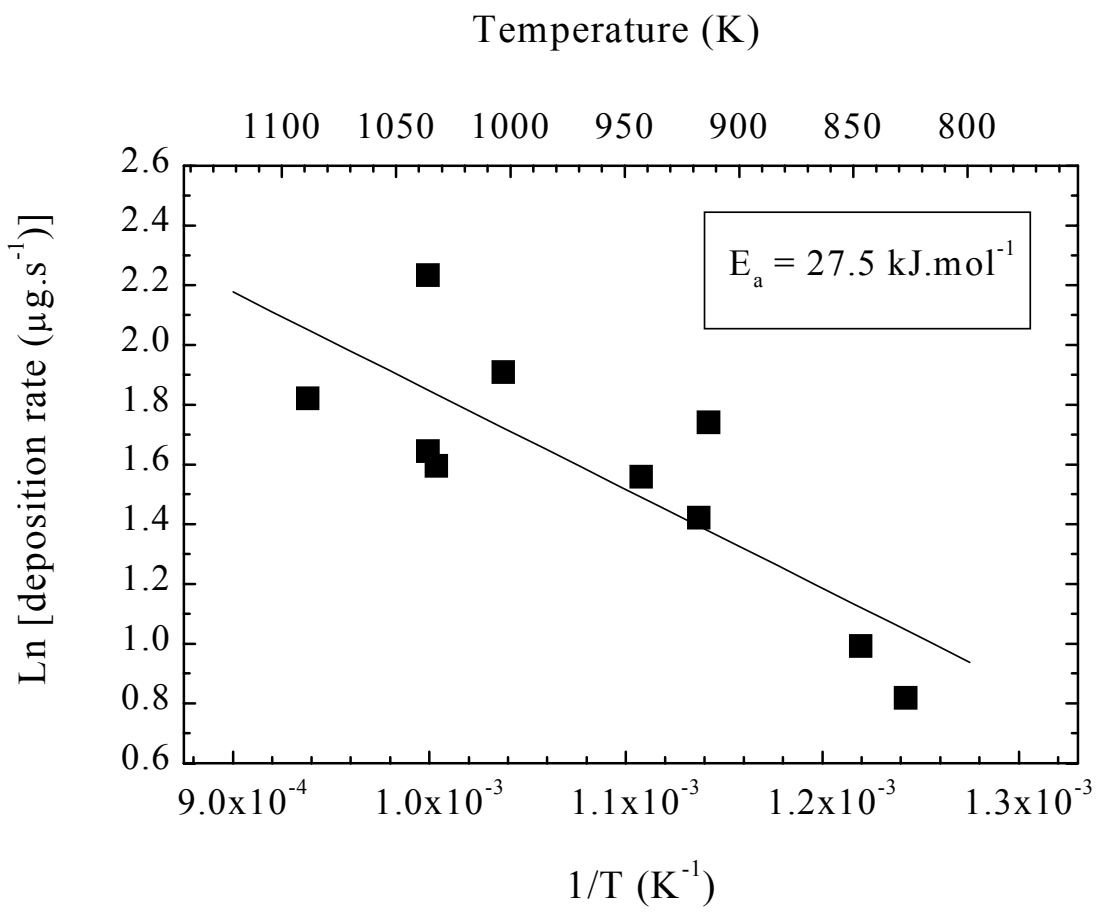

**Figure 2** - Logarithm of the apparent deposition rate as a function of reciprocal temperature. ■, measured values; —, curve fitting.

*Structural analysis*

The GIXRD diffractograms show narrow diffraction lines which are all matched by the rhombohedral boron carbide JCPDS card no. 33-0225. This result confirms that only one polycrystalline phase was formed during the LCVD process. Figure 3 displays a series of spectra recorded over samples with varying carbon content, from 9.8 to 18.8 at%, and also a standard bulk sample with 20 at% C, for comparison. The selected 2θ region is the one comprising the two strongest boron carbide diffraction peaks, i. e. the (104) and (021) reflections. When carbon concentration is high (fig. 3a and 3b) the experimental intensity distribution pattern follows the relative intensities of the standard polycrystalline $B_4C$ sample with grains randomly oriented. For films with low carbon contents an inversion of the relative intensities of the (104) and (021) peaks was always observed, suggesting the development of a (104) texture (fig. 3c-e). A strong continuous shift of the (021) peak to lower angles is also observed when the atomic carbon concentration decreases.

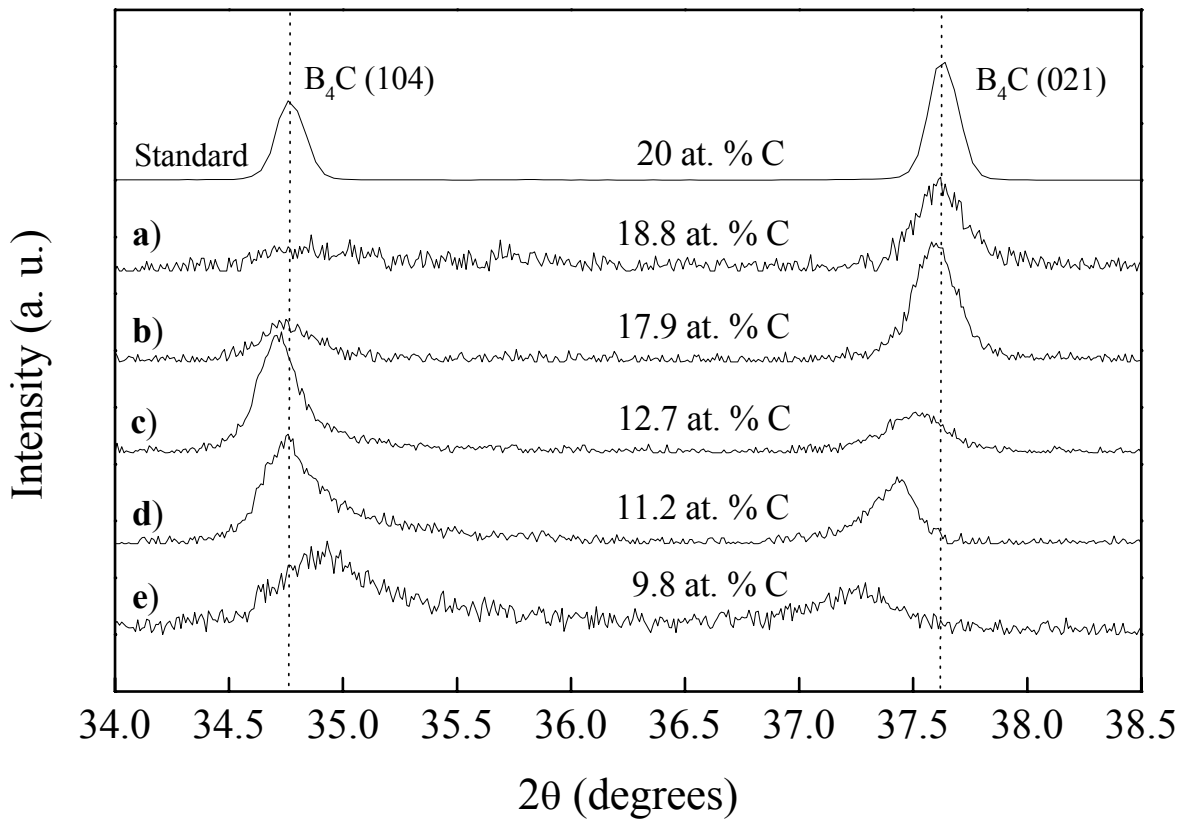

**Figure 3** - GIXRD spectra of boron carbide films with different carbon contents, as measured by EPMA.

Based on the $d_{hkl}$ values inferred from the angular position of the two major peaks in the GIXRD patterns, we estimated the $B_4C$ equivalent hexagonal lattice parameters $c_h$ and $a_h$ using the following equations

$$c_h = \sqrt{\frac{63\, d_{104}^2\, d_{021}^2}{4\, d_{021}^2 - d_{104}^2}} \quad ; \quad a_h = \sqrt{\frac{4\, c_h^2\, d_{104}^2}{3\,(c_h^2 - 16\, d_{104}^2)}}$$

(1)

which were deduced from the spacing formulae given by B. E. Warren [11]. The $c_h$ and $a_h$ constants are correlated with the $B_4C$ rhombohedral lattice parameters $a_r$ and $\alpha_r$ by the relations [11].

$$a_r = \frac{1}{3}\sqrt{3a_h^2 + c_h^2} \quad ; \quad \sin\frac{\alpha_r}{2} = \frac{3}{2\sqrt{3 + (c_h / a_h)^2}}$$

(2)

Figure 4 shows $c_h$ and $a_h$ parameters as a function of the carbon content of the films. The non-linear correlation found between the lattice constants and the carbon concentration suggests the analysis of the whole graph in terms of three distinct zones, which can be related to the way carbon atoms are preferentially substituted in the lattice as the carbon content is reduced. Between 20 and about 17.5 at% C (zone I) both $c_h$ and $a_h$ increase when the carbon concentration decreases. This behaviour may be explained by the exchange of carbon atoms by boron atoms in the central C-B-C chain leading to a C-B-B chain in the main diagonal of the rhombohedral structure. Since boron atom has a higher covalent radius (0.82 Å) than carbon atom (0.77 Å), these substitutions induce an increase on the volume of the hexagonal lattice, expanding the unit cell in both a and c dimensions.

Between about 17.5 and 13.5 at% C (zone II) the parameters $c_h$ and $a_h$ present values almost constant. This zone could be interpreted on the basis of substitutions of carbon atoms by boron atoms in the $B_{11}C$ icosahedral units since this kind of substitution does not affect significantly the volume of the unit cell. For carbon contents lower than 13.5 % (zone III) a decrease in carbon concentration yields a decrease on $c_h$ constant and an increase on $a_h$ constant.

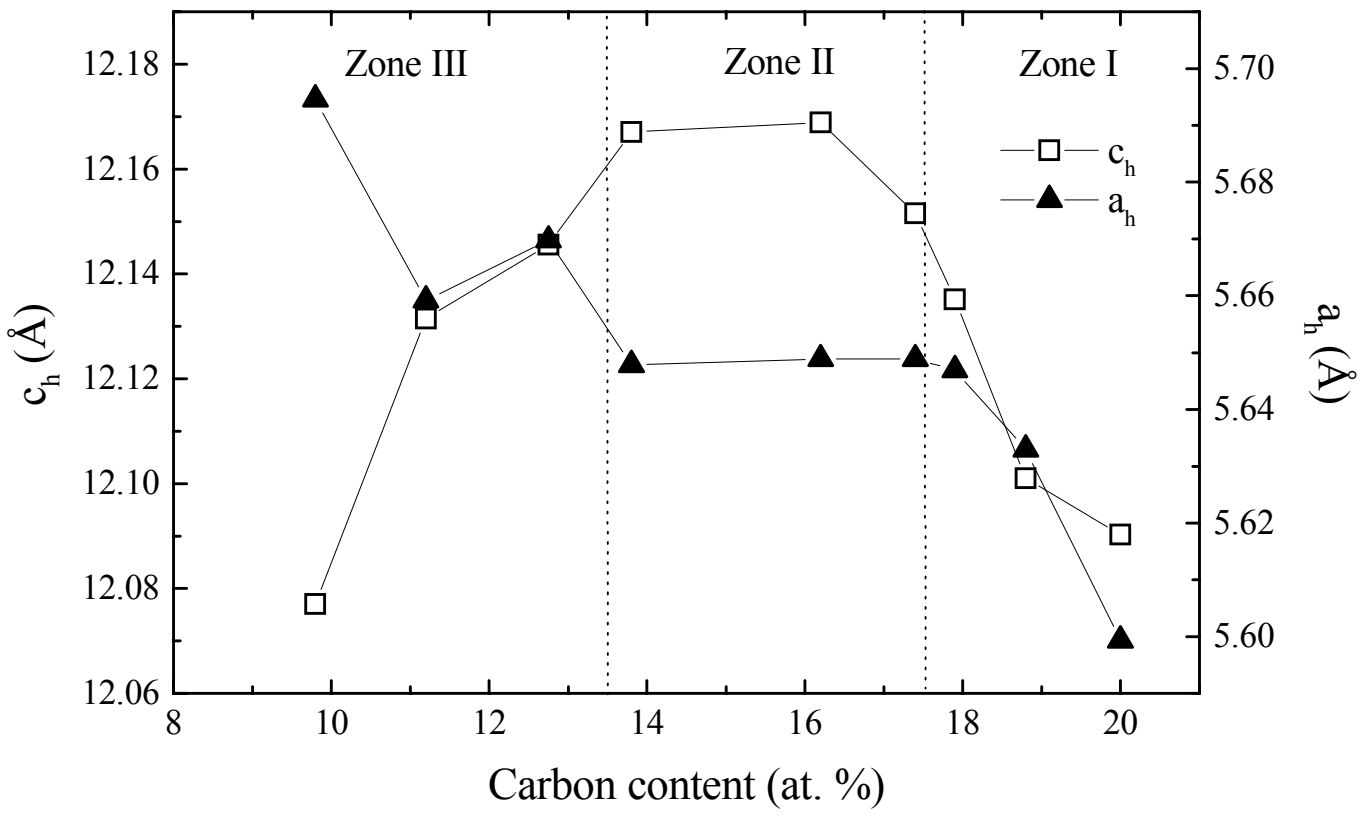

**Figure 4** - Hexagonal lattice parameters, $c_h$ and $a_h$, as a function of carbon content of the boron carbide films.

The results found in zone I and zone II and the intrinsic carbon $\rightarrow$ boron substitution sequence agree qualitatively with the evolution of the lattice parameters referred by Aselage et al. [7]. However, these authors extend zone I to about 13.5 at% C and zone II to about 9 at% C. Zone III is absent in the results of Aselage, probably due to the discrepancy in the measured carbon concentration. When converted to the rhombohedral lattice parameters, the measured values of $c_h$ and $a_h$ in zone III follow the trend reported by H. L. Yakel [12] for low carbon concentrations. Unlike the trends observed in zones I and II, the behaviour in zone III seems to be hardly explained considering only the prolonged replacement of carbon atoms by boron atoms on the main diagonal and/or on the icosahedral units. Therefore, one may attempt to interpret this stage of further carbon depletion by considering the replacement of the C-B-C and/or C-B-B chains by a new intericosahedra linking component, whose existence was already proposed by Yakel [12] and seems to be consistent with our micro-Raman results as will be seen in the following discussion.

Figure 5 shows a series of micro-Raman spectra obtained in the centre region of different boron carbide films. All the Raman spectra were normalised to the broad peak at 700 $cm^{-1}$ and the background was removed by empirical procedures. As a reference, fig. 5 also displays the micro--Raman spectrum obtained with a bulk standard sample of $B_4C$ with 20 at% C. The spectrum of the sample with 18.8 at% C (fig. 5a) presents in the high frequency region three broad peaks at 720, 840 and 1070 $cm^{-1}$. The broad peak at 1070 $cm^{-1}$ has a clear developed shoulder between 900 and 1000 $cm^{-1}$, which can be fitted by two peaks at 930 and 980 $cm^{-1}$. These peaks match the peaks assigned by Tallant et al. [6] to the breathing modes of the icosahedral $B_{11}C$ structures. In the low frequency region spectrum 5a presents three narrower peaks at 320, 480 and 530 $cm^{-1}$. The peak at 320 $cm^{-1}$ is an experimental artefact which appears in all Raman spectra. The other two peaks at 480 and 530 $cm^{-1}$ were also reported by Tallant as due to the vibration modes of the C-B-C chains. By comparing this micro-Raman spectrum (fig. 5a) with the one of the standard sample (20 at% C) it can be observed that the Raman bands between 600 and 1200 $cm^{-1}$, which are associated with the icosahedral modes, are relatively unaffected by the decrease of carbon content in the samples. However, in the low frequency region assigned to the vibrations of C-B-C chains, the doublet at 480/530 $cm^{-1}$ decreases in intensity while a new peak at 380 $cm^{-1}$ tends to develop. These results are consistent with the replacement, during the initial stages of carbon depletion, of the C-B-C chains along the main diagonal by C-B-B chains, and support the behaviour of the lattice parameters found in zone I of figure 4. For the same type of vibrational modes, the less stiff C-B-B chains will have a lower vibration frequency [6,7].

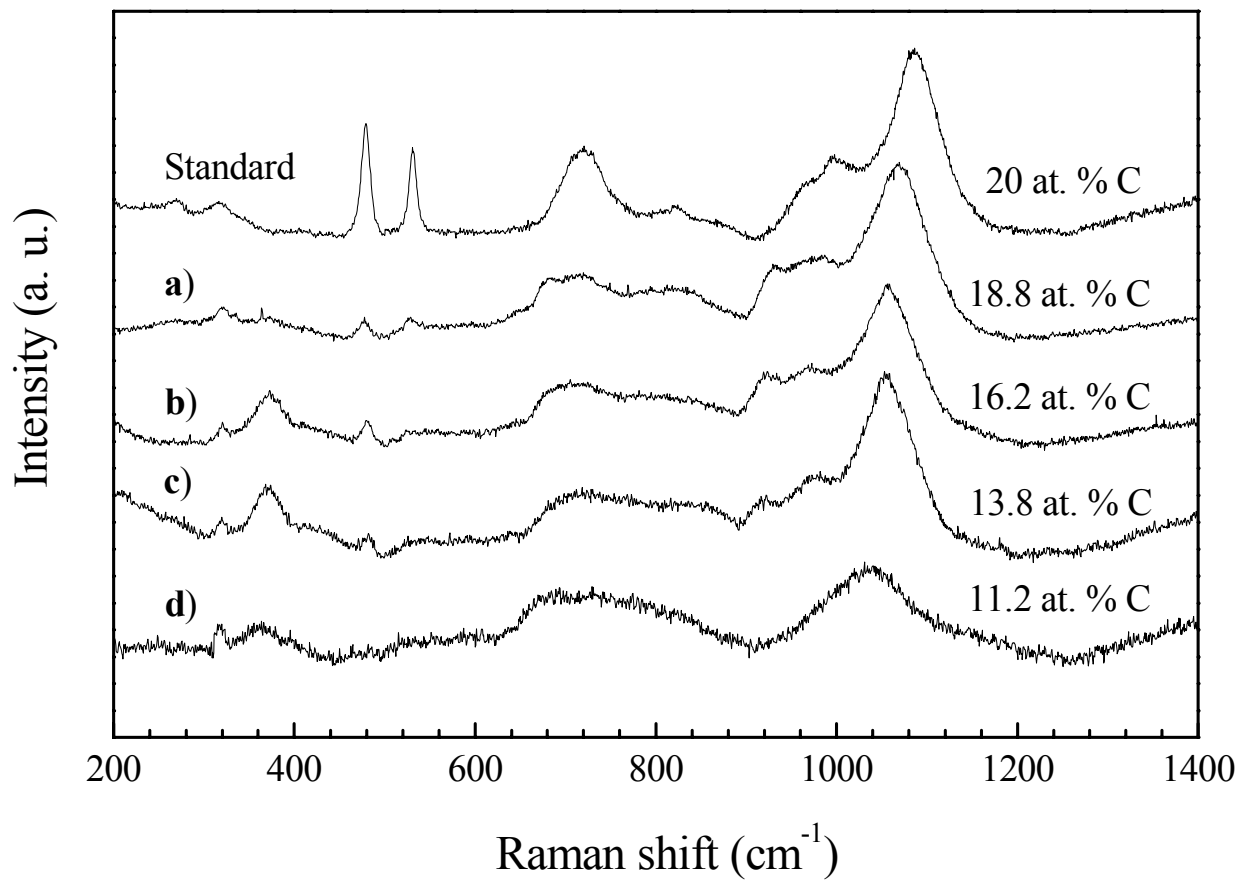

**Figure 5** - Raman spectra of boron carbide films with different carbon contents, as measured by EPMA.

The micro-Raman spectra of figures 5b and 5c concern films with carbon concentration values of 16.2 and 13.8 at%, respectively. The carbon content of these two samples corresponds to zone II of figure 4. By comparing both spectra with spectrum 5a we can observe that the peaks at 480 and 530 $cm^{-1}$ become less intense while the 380 $cm^{-1}$ is enhanced as the carbon content decreases. Also, in the high frequency range, the 1070 $cm^{-1}$ peak is shifted to lower frequency, from 1070 to 1055 $cm^{-1}$, while the intensity of the shoulder of this peak also decreases. These results indicate that the icosahedral $B_{11}C$ units are also affected by the decreasing amount of carbon, leading to $B_{12}$ icosahedra [6,7]. By comparing the spectra 5b with 5c, both related to samples inside zone II of fig. 4, but with decreasing carbon content, the most affected bands are clearly the high frequency ones. Therefore, it can be concluded that the most probable carbon $\rightarrow$ boron substitution in zone II occurs in the $B_{11}C$ icosahedra.

Figure 5d was recorded over the centre region of a film with carbon concentration of 11.2 at%, corresponding to zone III of figure 4. By analysing this spectrum and comparing it with the ones described above, we can conclude that a further decrease in carbon content leads to a substantial modification of the micro-Raman spectra of boron carbide samples. In the high frequency region, the peaks at 1055 $cm^{-1}$ become broader, less intense and shift to lower frequency by about 15 $cm^{-1}$. In the low frequency region the peaks at 480 and 530 $cm^{-1}$ vanish and the intensity of the broad peak at 380 $cm^{-1}$ strongly decreases, eventually disappearing at lower carbon concentration. These Raman results are a clear indication that, for carbon atomic concentration in zone III, the C-B-C and/or C-B-B chains are strongly affected, and can even disappear at low carbon content values. This seems to be consistent with the hypothesis of Yakel [12], i.e. the randomly replacement of the main diagonal chains by a new linking component consisting of four boron atoms located at the centre of the unit cell in a plane perpendicular to the $c_h$ axis. As a result of the substitution of the C-B-C and/or C-B-B chains by these ($B_4$) planar groups, Yakel anticipated an increase in the rhombohedral unit cell angle ($\alpha_r$) with a little effect on cell edge ($a_r$). Our results are in good agreement with these predicted trends since after converting the hexagonal lattice parameters into rhombohedral lattice parameters by using relations (2), the values found in zone III show an increase of $\alpha_r$ of about 2 % while the $a_r$ values are almost constant, with a fluctuation less than 0.15 % when the carbon content decreases from 13.5 to 9.8 at%.

## Conclusions

Boron carbide films were deposited on fused silica substrates by $CO_2$ laser CVD, using a gas mixture of $BCl_3$, $CH_4$, $H_2$ and Ar. The deposited films present good adherence, a fine grain morphology and a mean carbon concentration in the range from 9 to about 20 at%. Growth rates in the range 2.3 – 9.4 $\mu g.s^{-1}$ were deduced from the thickness profiles of the films. The Arrhenius equation for the deposition rate yields an apparent activation energy of 27.5±7.8 $kJ.mol^{-1}$, showing that for the experimental conditions used in this study, the mass transport of reactive gaseous species is the rate-limiting step for the boron carbide film growth.
The crystallographic lattice constants, $c_h$ and $a_h$, were plotted as a function of the carbon content in the films, showing a non-linear dependence for these two quantities. The trends found are consistent with the micro-Raman spectra of the samples and allow to interpret the way in which carbon atoms are preferentially substituted by boron atoms in the lattice as the carbon content is reduced. As the carbon concentration starts to decrease, C-B-C chains are the units preferentially affected, by replacement by C-B-B chains. Between about 17.5 and 13.5 at% C, the most affected structures are the $B_{11}C$ icosahedra, leading to the formation of $B_{12}$ structure units. Bellow 13.5 at% C, the C-B-C and/or C-B-B diagonal chains are affected and probably disappear at low carbon concentrations.

## Acknowledgements

This work was partially funded by EU under contract BRE2-CT93-0451. J.C. Oliveira gratefully acknowledge MSc research grant from JNICT (P).